\documentclass[11pt]{article}
\usepackage{epsf,amsmath,amssymb}

\newenvironment{mydescription}[1]%
 {\begin{list}{}{%
  \setlength{\leftmargin}{10pt}%
  \setlength{\itemindent}{10pt}%
  \setlength{\rightmargin}{10pt}}}
  {\end{list}}

\begin{document}

\begin{flushright}
 ITFA 98-24  \\
 BI/TP-98/17 \\
 October 1998
\end{flushright}

\begin{center}

\vspace{24pt}

{\Large \bf Simulating Four-Dimensional Simplicial Gravity
 \vspace{5pt}
  using Degenerate Triangulations}
\vspace{24pt}

{\large \sl Sven~Bilke $^{a}$ \,{\rm and}\, Gudmar Thorleifsson $^{b}$ } \\
\vspace{10pt}
$^a$ Inst.\ Theor.\ Fysica, Univ.\ Amsterdam, 1018 XE Amsterdam,
     The Netherlands \\
$^b$ Facult\"{a}t f\"{u}r Physik, Universit\"{a}t Bielefeld
     D-33615, Bielefeld, Germany  \\
\vspace{10pt}

\begin{abstract}
We extend a model of four-dimensional simplicial quantum
gravity to include {\it degenerate} triangulations in addition
to combinatorial triangulations traditionally used.
Relaxing the constraint that every 4-simplex is uniquely
defined by a set of five distinct vertexes, 
we allow triangulations containing multiply connected simplexes 
and distinct simplexes defined by the same set of vertexes. 
We demonstrate numerically that including degenerated 
triangulations substantially reduces 
the finite-size effects in the model.  
In particular, we provide a strong numerical evidence for 
an exponential bound on the entropic growth of the
ensemble of degenerate triangulations, and show that a
discontinuous crumpling transition is already observed on 
triangulations of volume $N_4 \approx 4000$.
\\
\vspace{5pt}
\noindent
PACS numbers: 04.60.-m,Nc \quad 05.70.Fh \quad 02.70.Lq
\end{abstract}

\end{center}
\vspace{15pt}

Discretized models of four-dimensional Euclidean quantum gravity, 
known as simplicial gravity or dynamical triangulations,
have received ample attention in recent years,
the hope being that in a suitable scaling limit
they might provide a sensible non-perturbative definition
of quantum gravity.  
In the simplicial gravity approach the integration over metrics is replaced 
by a sum over an ensemble of triangulations constructed
by all possible gluings of equilateral 4-simplexes into 
closed (piece-wise linear) simplicial manifolds 
(see e.g.\ Ref.~\cite{david,janbook}).  
The regularized Euclidean Einstein-Hilbert action is 
particularly simple; it can be taken to depend on
only two coupling constants, $\mu$ and $\kappa$,
related to the cosmological and the inverse Newton's 
constants.  The coupling constants
are conjugate to the volume --- the number of 4-simplexes --- 
and the number of triangles in a given triangulation respectively.
The regularized grand-canonical partition function thus becomes:
\begin{equation}
Z(\mu,\kappa) \;=\; \sum_{T\in{\cal T}} \; \frac{1}{C_T} \;
{\rm e}^{\textstyle -\mu N_4 + \kappa N_2}.
\label{model}
\end{equation}
The sum is over all distinct triangulations $T \in {\cal T}$,
$N_i$ is the number of $i$-simplexes in a
triangulation $T$ and $C_T$ denotes its
symmetry factor --- the number of
equivalent labeling of the vertexes.

Extensive numerical simulations 
have established that the model Eq.~(\ref{model}) 
has a strong-coupling (small $\kappa$) crumpled phase
and a weak-coupling (large $\kappa$) elongated phase,
separated by a discontinuous phase transition.  
In the crumpled phase the internal geometry collapses and
is dominated by a novel {\it singular} structure --- two
singular vertexes connected to an extensive fraction of the 
total volume and joined by a sub-singular edge \cite{singular}.
The elongated phase, on the other hand, is dominated by 
branched polymer like triangulations,
i.e.\ bubbles glued together {\it via} small
necks into a tree-like structure. 

In Eq.~(\ref{model}), ${\cal T}$ denotes a suitable ensemble of 
triangulations included in the partition function.
Different ensembles are defined by imposing various
restrictions on how the simplexes are glued together.  
Provided this leads to a well-defined partition function,
and as long as the difference is only at the level
of discretization, one expects different choices
of ${\cal T}$ to lead to the same continuum theory
in the thermodynamic limit.  This is known to be true in two
dimensions where models of simplicial gravity 
corresponding to different choices of ${\cal T}$
are soluble as matrix models \cite{mat2d}.
Even in two dimensions, however, for the
partition function to be convergent the topology
of the triangulations included in the sum Eq.~(\ref{model}) 
must be fixed, regardless of the ensemble used.
As the same most likely is true in higher dimensions,
in this letter we only consider triangulations of fixed 
spherical topology.

All simulations of four-dimensional simplicial gravity have,
as of yet, used an ensemble  of {\it combinatorial} triangulations 
${\cal T}_C$.  In a combinatorial triangulation every 
${\rm D}$-simplex is uniquely defined by a set of 
(${\rm D}+1$) distinct vertexes --- it is said
to be combinatorially unique.  In this letter we study
a larger ensemble of {\it degenerate} triangulations ${\cal T}_D$
where we relax this constraint and allow distinct simplexes
to be defined by the same set of vertexes.  This includes
two simplexes with more than one face in common.   
We do, however, retain
the restriction that every 4-simplex is defined by a set
of five distinct vertexes, i.e.\ we exclude degenerate simplexes.
Clearly ${\cal T}_C \subset {\cal T}_D$.
This corresponds to an ensemble of restricted
degenerate triangulations as defined in Ref.~\cite{deg3d}.

The benefits of using a larger ensemble of 
triangulations are well known from simulations of
two-dimensional simplicial gravity which have
demonstrated that less restricted the
triangulations are translates into smaller 
finite-size effects \cite{fss2d}.  
Recently the same observation
has been made in three dimensions \cite{deg3d}.  
As simulations of four-dimensional
simplicial gravity are notoriously time-consuming, 
primarily due to the large volumes needed to
observe any ``true'' infinite-volume behavior, any 
reduction in the finite-size effects 
is of great practical importance.

In this letter we show that including degenerates
triangulations in simulations of four-dimensional
simplicial gravity likewise leads to reduced finite-size 
effects. This reduction is most pronounced in the crumpled phase
of the model where the free-energy density of the canonical 
(fixed volume) ensemble --- the pseudo-critical cosmological 
constant $\mu_c(N_4)$ --- converges very rapidly to an 
infinite-volume value.  This in turn implies an exponential bound on 
the entropic growth of the ensemble of degenerate triangulations, 
something that has been the subject of some controversy 
in the past for combinatorial triangulations \cite{bound4d,bound2}.
Although we observe qualitatively the same phase structure as
with the model Eq.~(\ref{model}) restricted to combinatorial
triangulations, there are some dissimilarities.
A discontinuous phase transition, separating the elongated and 
the crumpled phases, is already observed on triangulations of relatively 
modest size, $N_4 \approx 4000$, 
compared to combinatorial triangulations were a volume
of $N_4 \approx 32.000$ is needed.
And while the crumpled phase is still dominated by a singular
structure, for degenerate triangulations this 
corresponds to a {\it gas} of sub-singular vertexes rather than
to only two singular vertexes.

\vspace{5pt}

We have simulated the model Eq.~(\ref{model}) using 
degenerate triangulations on volumes up to 32.000 4-simplexes
using Monte Carlo methods.  As customary
we work in a quasi-canonical ensemble of 
spherical manifolds with almost fixed $N_4$:
\begin{equation}
 Z(\mu,\kappa;\bar{N}_4) \;=\; \sum_{N_4} 
  {\rm e}^{\textstyle \;-\mu N_4 
   -\delta (N_4 - \bar{N}_4)^2 } \; \Omega_{N_4}(\kappa),
 \label{pseudo}
\end{equation}
where 
$\Omega_{N_4}(\kappa) = \sum_{T\in{\cal T}(N_4)}\exp (\kappa N_2)$ 
is the canonical partition function.  
As there do not exist ergodic volume conserving
local moves, hence the canonical ensemble cannot be simulated
directly, we must allow the volume to fluctuate.
The quadratic potential term added to the action ensures, 
for an appropriate choice of $\delta$, that these fluctuations
are small.
 
In the simulations the triangulation space
is explored using a set of local geometric changes, 
the ($p,q$)-moves.
In a ($p,q$)-move, where $p = {\rm D}+1-q$,  a ($q-$1)--subsimplex in
the triangulation is replaced by its ''dual`` ($p-$1)--subsimplex. 
For combinatorial triangulations the ($p,q$)--moves are known to be 
ergodic for ${\rm D} \leq 4$ \cite{moves}.  
To demonstrate that the same holds true for degenerate triangulations
we observe that, just as in three dimensions \cite{deg3d}, every
set of combinatorially equivalent simplexes, or sub-simplexes, 
can be made distinct by a finite sequence of the ($p,q$)--moves.
Thus every degenerate triangulations can be reduced to a
combinatorial one.  In addition the
local nature of the ($p,q$)--moves prohibits the creation of 
pseudo-manifolds in the simulations, i.e.\ triangulations 
containing vertexes with a 
neighborhood not homeomorphic to the ${\rm D}$-ball.

From a practical point of view simulating degenerate triangulations
is actually simpler than simulating their combinatorial counterpart as
one avoids the non-local manifold checks necessary to
exclude combinatorially equivalent simplexes. For combinatorial
triangulations these checks are the most time-consuming part
of the simulations \cite{simon}.  This simplification of
is particularly beneficial in
the crumpled phase where the singular structure dominates;
in this phase we observe a tenfold reduction in the effective 
auto-correlation times (measured in ``real'' time) when
using degenerate instead of combinatorial triangulations.  
In the branched polymer phase, on the other hand, 
the auto-correlation times appear
comparable for the two ensembles.

\begin{figure}[t]
\epsfxsize=4.2in \centerline{\epsfbox{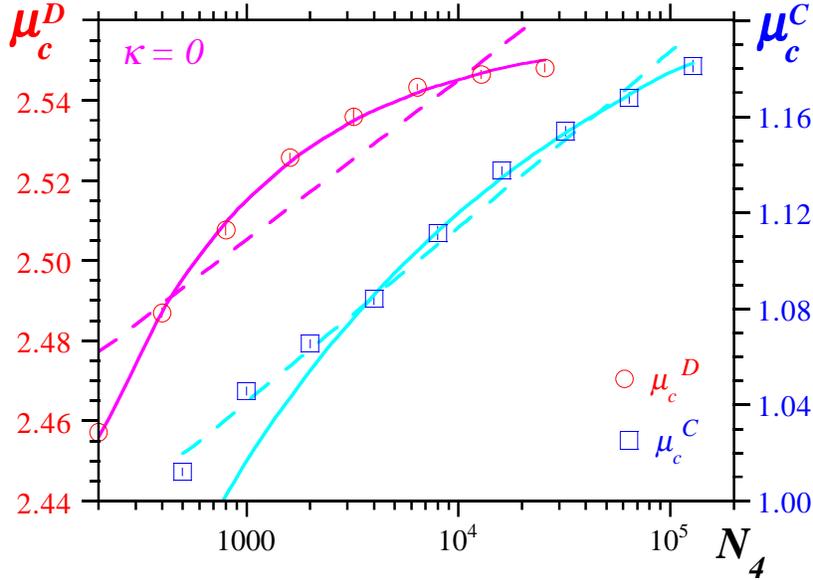}}
\caption{\small The pseudo-critical cosmological constant 
 $\mu_c^D(N_4)$ for an ensemble of degenerate 
 triangulations, together with fits assuming a power-law convergence,
 Eq.~(\ref{exp}) (solid curve), 
 or a logarithmic divergence, Eq.~(\ref{super}) (dashed curve).
 Also included are the corresponding values, $\mu_c^C(N_4)$, for
 an ensemble of combinatorial triangulations.}
\label{fig1}
\end{figure}

\begin{table}
\caption{\small The parameters in the fit of the
 pseudo-critical cosmological constant $\mu_c^D(N_4)$
 to Eqs.~(\ref{exp}) and (\ref{super}) respectively.  Measurements 
 on volumes $N_4 = 400$ to 25.600 are included in the fits.}
 \begin{center}
  \begin{tabular}{|llll|}  \hline
 & $\bar{\mu}$ & $\gamma$ & $\chi^2$/d.o.f.    \\ \hline
 \vspace{-8pt} &&& \\ 
 Eq.~(\ref{exp})     & 2.556(3) & 0.55(5)  & [3.8] \\
 Eq.~(\ref{super})  & 2.385(4) &   & [117]  \\ 
  \vspace{-8pt} &&& \\ \hline 
\end{tabular}
\end{center}
\end{table}

The real benefit of using degenerate triangulations is
the reduction of geometric finite-size effects. 
This reduction is most striking for the volume dependence of the
pseudo-critical cosmological constant, $\mu_c^D(N_4)$, which
we shown in Figure~1.   
For comparison we also include the corresponding 
values, $\mu_c^C(N_4)$,  for combinatorial triangulations.
For degenerate triangulations we observe a rapid
convergence to an infinite volume value $\bar{\mu}$.
This can be quantified by comparing the fit of
$\mu_c^D(N_4)$ to two different functional
forms: a weak power-law convergence,
\begin{equation}
\mu_c(N_4) \;=\; \bar{\mu} + \frac{b}{N_4^{\gamma}},
\label{exp}
\end{equation}
or a logarithmic divergence,
\begin{equation}
\mu_c(N_4) \;=\; \bar{\mu} + b^{\prime} \log N_4.
\label{super}
\end{equation}
The fit parameters and the quality of the fits are shown in
Table~1.  In contrast to combinatorial triangulations
for degenerate triangulations there is no comparison in the
quality of the fits; the latter, which corresponds to a divergent
partition function Eq.~(\ref{model}), is ruled out by a 
$\chi^2/({\rm d.o.f.}) \approx 117$.  For combinatorial
triangulations, on the other hand, it is difficult to use the quality of 
the fit of $\mu_c^C(N_4)$ to either Eq.~(\ref{exp}) or Eq.~(\ref{super})
to distinguish between those two scenarios 
(see e.g.\ Ref.~\cite{bound2}).  

The importance of this result is that it
provides strong numerical evidence for an exponential 
bound on the entropic growth of the ensemble of degenerate triangulations
as a function of volume --- a necessary condition for a well-defined
partition function Eq~(\ref{model}).  
And, as ${\cal T}_C \in {\cal T}_D$, this 
implies an exponential bound on the number of combinatorial
triangulations as well. 

The origin of the large finite-size effects present
in simulations with combinatorial triangulations lies
in the nature of the quantum geometry in the crumpled phase.
As stated, the partition function is dominated by triangulations 
characterized by two singular vertexes connected 
by a sub-singular edge.  
A singular vertex has a local volume $q$
--- the number of 4-simplexes containing that vertex --- 
which grows linearly with the volume of the manifold, 
while the sub-singular edge
has a local volume which grows like $N_4^{\alpha}$, 
$\alpha \approx 2/3$ \cite{singular}.
However, for combinatorial triangulations this singular
structure only dominates on large enough volumes, 
on small volumes triangulations with only
one singular vertex have larger entropy.
This results in a cross-over behavior in the
fractal structure at $N_4 \approx 1000$, as can
be observed in Figure~1.

\begin{figure}[t]
 \epsfxsize=4.2in \centerline{\epsfbox{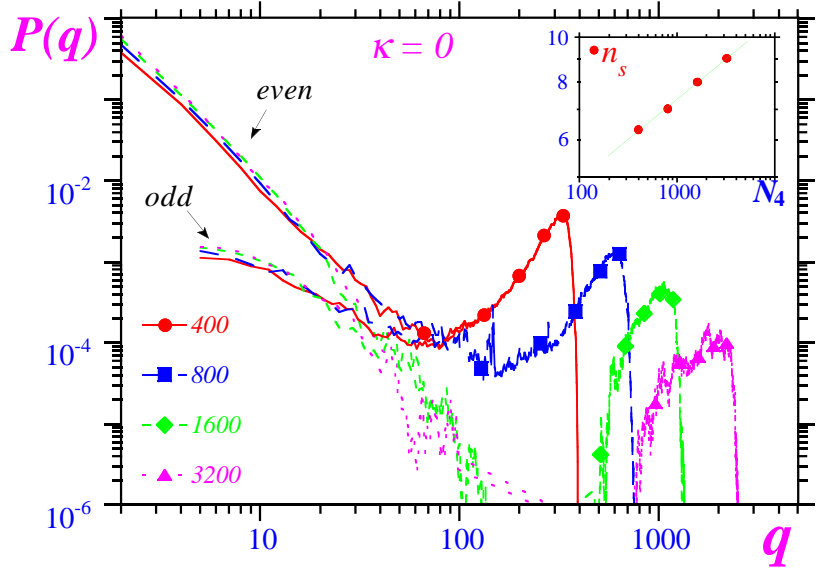}}
\caption{\small The (normalized) distributions of local volumes, 
 $P(q)$, for degenerate triangulations. This is for
 $400 \leq N_4 \leq 3200$, and for $\kappa = 0$.
 ({\sf Insert})  The number, $n_s$, of sub-singular vertexes
 {\em versus} volume.}
\label{fig2}
\end{figure}

For degenerate triangulation the crumpled phase is likewise
dominated by a singular structure.  This is evident from
the probability distribution $P(q)$ of the
local volumes which contains an isolated peak in the tail.  
This is shown in Figure~2 for $\kappa = 0$.   
However, the distribution $P(q)$ differs in two respects
from the corresponding distribution measured on
combinatorial triangulations \cite{singular}:

\begin{mydescription}

\vspace{-5pt}
\item[({\sf a)}]
 The peak corresponds to not just two singular
 vertexes but rather to several sub-singular vertexes,
 i.e.\ vertexes with local volumes that scale like 
 $N_4^{\alpha}$, $\alpha < 1$.  A rough estimate yields
 $\alpha \approx 0.9$. The number of these sub-singular vertexes,
 $n_s$, increases logarithmically with the volume
 as is shown in the insert in Figure~2.  This suggests that the
 crumpled phase is dominated by a {\em gas} of sub-singular
 vertexes.
 
\vspace{-5pt} 
\item[({\sf b)}]
 For each volume, $P(q)$ effectively separates into two 
 distinct distributions depending on whether the local volume
 is even or odd.
 
\end{mydescription}

\vspace{-5pt}
It is not clear though how much significance should
be attached to this difference in the singular structure.
Due to the collapsed nature of the internal geometry  
it is unlikely that any sensible continuum limit exists
in the crumpled phase, hence there is no reason
to expect identical scaling behavior for the two different ensembles.  
The details of the discretization may still be important in the 
thermodynamic limit for $\kappa < \kappa_c$.
It is, however, worth noticing that for degenerate triangulations
we do not observe any change in the fractal structure as
the volume is increased as for combinatorial triangulations.

Additional evidence of a collapsed intrinsic geometry in
the crumpled phase
comes from the (absence of) volume scaling of the simplex-simplex
distribution $s(r)$, i.e.\ the number of simplexes 
at a geodesic distance $r$ from a marked simplex.   
Using the scaling ansatz $s(r) = N_4^{1-1/d_H} F(x)$,
where $x = r/N_4^{1/d_H}$ \cite{dhaus,jands},
we tried to collapse distributions $s(r)$ measured on
different volumes onto a single scaling curve.  
This was though not possible with an ``acceptable''
collapse (with $\chi^2/({\rm d.o.f.})$ of order unity); moreover,
the estimate of the fractal dimension $d_H$ appeared to increase 
with the volume. From this we conclude that $d_H = \infty$ 
in the crumpled phase.

\begin{figure}[t]
 \epsfxsize=4.2in \centerline{\epsfbox{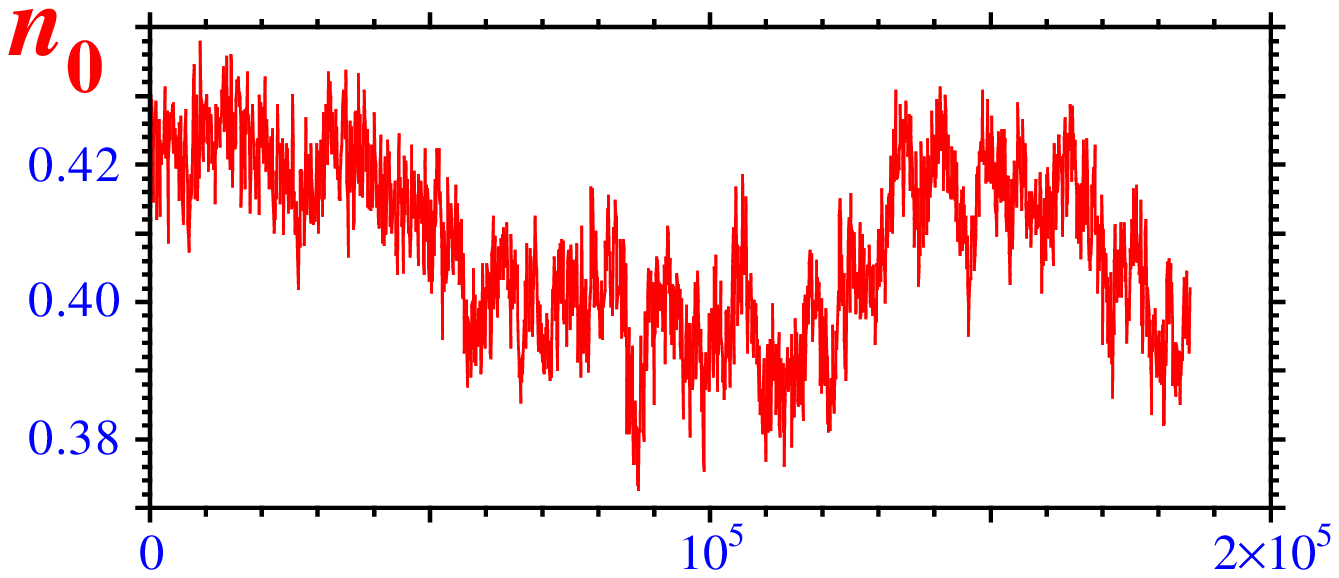}}
 \epsfxsize=4.2in \centerline{\epsfbox{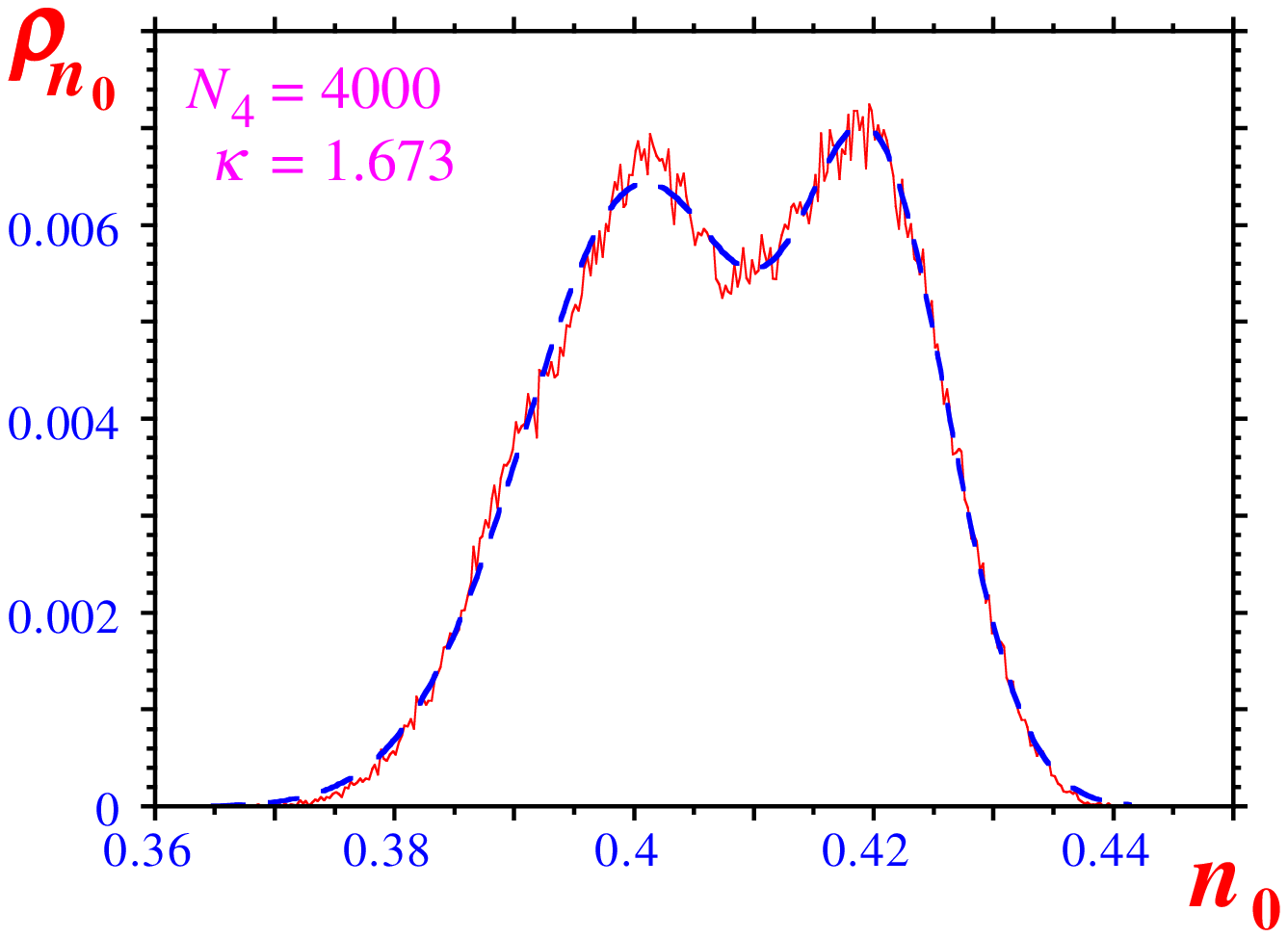}}
\caption{\small ({\sf Top}) A MC time-series of the energy density,
 $n_0 = N_0/N_4$, for $N_4=4000$ and $\kappa = 1.673$.
 ({\sf Bottom})  The corresponding histogram together with a
  fit to a function composed of two Gaussian peaks (dashed line).}
\label{fig4}
\end{figure}

\vspace{5pt}
We have also investigated the phase structure of
the model for non-zero values of the inverse Newton's constant
$\kappa$.  As for combinatorial triangulations we observe
a phase transition to an elongated phase at  
$\kappa_c \approx 1.7$.  
To establish the nature of the phase transition 
we have studied the Monte Carlo time-series of the 
energy density, $n_0 = N_0/N_4$, in the critical region.
We show an example of one such time-series in Figure~3 together
with the corresponding histogram.
This is for volume $N_4 = 4000$ and $\kappa = 1.673$.  
The histogram shows a clear double-peak
structure characteristic of a discontinuous phase transition.  
Note that in order to observe the corresponding two-state
signal using the combinatorial ensemble, triangulations of
volume $N_4 \approx 32.000$ are needed \cite{firstord}. 

For $\kappa > \kappa_c$ the model is in an elongated
or branched polymer phase.  This we have established by
measuring the fractal dimensions $d_H$ and the spectral
dimension $d_s$ for $\kappa = 2$.  The former is determined
from the scaling of the simplex-simplex distribution, the latter
from the return probability of a walker on the dual graph,
$p(t) \sim t^{-d_s/2}$ \cite{jands}.  Including measurements on volumes
$N_4 = 400$ to 1600, we get $d_H =1.9(2)$ and $d_s = 1.32(5)$,
in excellent agreement with $d_H = 2$ and $d_s = 4/3$ as
expected for branched polymer.

\vspace{5pt}
In this letter we have demonstrated that 
including degenerate triangulations in simulations of four-dimensional
simplicial gravity has many potential advantages over the
model restricted to combinatorial triangulations.  This agrees
with the same observations previously made in both two
and three dimensions.  The chief benefit is the reduction
in geometric finite-size effects manly due to an
enlarged ensemble --- with a larger triangulation-space      the
infinite-volume fractal structure is more easily approximated
on small volumes.  
The most important result presented in this letter
is a strong numerical evidence for an
exponential bound on the entropic growth of
the canonical partition function $\Omega_{N_4}$.
A more practical result is the observation of
a discontinuous phase transition on triangulations of relatively
modest size. 

A natural extension of the work presented in this letter is
to investigate how the phase structure is affected 
if the model Eq.~(\ref{model}) is changed either by using a modified
measure \cite{bm} or by adding matter fields \cite{mat}.  
For combinatorial triangulations
it has recently been observed that this can substantially alter
the phase structure and, for a suitable modification,
a new {\it crinkled} phase appears \cite{us2}.  
If this observed phase structure corresponds to a genuine
change in the continuum behavior of the model Eq~(\ref{model}), 
one expects on basis of universality that it should be independent of
the ensemble of triangulations used.

\vspace{5pt}
\noindent
{\bf Acknowledgments:} 

We are indebted to  A.~Krzywicki and B.~Petersson for discussions.
S.B.\ was supported by FOM and G.T.\ by the Humboldt Foundation.

\end{document}